\documentclass[preprint,showpacs,showkeys,amsmath,amssymb,aps,doublespace]{revtex4}
\usepackage[dvips]{graphicx}
\usepackage{setspace}
\usepackage{amsmath}
\usepackage{amsfonts}
\usepackage{amssymb,color}
\usepackage{graphicx,color}
\usepackage[latin1]{inputenc}

\begin{document}

\title{Phase diagram of the symbiotic two-species contact process}

\author{Marcelo Martins de Oliveira$^1$\footnote{email: mmdeoliveira@ufsj.edu.br}
and Ronald Dickman$^2$\footnote{email: dickman@fisica.ufmg.br}
}
\address{
$^1$Departamento de F\'{\i}sica e Matem\'atica,
CAP, Universidade Federal de S\~ao Jo\~ao del Rei,
36420-000 Ouro Branco, Minas Gerais - Brazil\\
$^2$Departamento de F\'{\i}sica and
National Institute of Science and Technology for Complex Systems,
ICEx, Universidade Federal de Minas Gerais,
C. P. 702, 30123-970 Belo Horizonte, Minas Gerais - Brazil
}

\date{\today}

\begin{abstract}

We study the two-species symbiotic contact process (2SCP), recently proposed in
[de Oliveira, Santos and Dickman, Phys. Rev. E {\bf 86}, 011121 (2012)] .
In this model, each site of a lattice may be vacant or host single individuals of
species A and/or B. Individuals at sites with both species
present interact in a symbiotic manner, having a reduced death rate, $\mu < 1$.
Otherwise, the dynamics follows the rules of the basic CP, with individuals
reproducing to vacant neighbor sites at rate $\lambda$
and dying at a rate of unity. We determine the full phase diagram in the $\lambda-\mu$ plane
in one and two dimensions by means of exact numerical quasistationary distributions, cluster approximations,
and Monte Carlo simulations.
We also study the effects of asymmetric creation rates and diffusion of individuals.
In two dimensions, for sufficiently strong symbiosis (i.e., small $\mu$),
the absorbing-state phase transition becomes
discontinuous for diffusion rates $D$ within a certain range.
We report preliminary results on the critical surface and tricritical
line in the $\lambda-\mu-D$ space.
Our results raise the possibility that strongly symbiotic associations of mobile species
may be vulnerable to sudden extinction under increasingly adverse conditions.

%}

\end{abstract}


\maketitle

\section{Introduction}

Originally proposed as a toy model for epidemic spreading, the contact process (CP) \cite{harris-CP}
can also be interpreted as
a stochastic single species birth-and-death process with a spatial structure \cite{durrett}.
In the CP, each individual can reproduce assexually with rate $\lambda$, or die with unitary rate.
When the reproduction rate $\lambda$ is varied, the system undergoes a phase
transition between extinction and survival.

Interacting, spatially extended, multi-species processes are a subject of recent interest
\cite{jansen,iwata,multisp,tauber,tubay,parasites,competing}. In particular, multispecies (or multitype)
contact processes have been used to model systems with neutral community structure, and have
proven useful in understanding abundance distributions and species-area relationships \cite{weitz, munoz}.

Symbiosis is the ``living together of two phylogenetically unrelated species in close
association" \cite{boucher}, and is thought to develop as a consequence of coevolution \cite{douglas,sapp};
it is a rather common phenomenon in nature. For example, lichens are symbiotic complexes of algae living inside fungi,
and the roots of higher plants use symbiotic associations with fungi to receive important nutrients \cite{paracer}.

Macroscopic models derived from modifications of the Lotka-Volterra competition equations
have been employed to model symbiotic relations for decades \cite{rockwood,yukalov}. Such model however neglect
stochastic effects, relevant due to the discrete nature of the individuals
and in spatially extended systems \cite{discrete}. More recently, the effects of mutualistic interactions
in one-dimensional stepping stone models
were studied by Korolev and Nelson \cite{korolev}, and by  Dall'Asta et.al. \cite{asta}, who found
that fluctuations and spatial structure favors symmetric mutualism (in which species benefit equally from the
interaction). The fixation(absorbing)-coexistence(active) phase transition was found to belong
to the voter model universality class if mutualism is symmetric, and to the directed percolation class if asymmetric.
Lavrentovich and Nelson extended the results of \cite{lav} to asymmetric interactions in two and three dimensions,
finding that the mutualist phase is more accessible in higher dimensional range expansions.
Pigolotti et. al \cite{pigolotti} studied competition and cooperation between two species
when the population size is not constrained as it is in stepping-stone models. 

Recently, we studied symbiotic interactions in a two-species CP \cite{scp}.  This was done by allowing two CPs
(species A and B), to inhabit the same lattice. The symbiotic interaction is modeled via a reduced death rate, $\mu < 1$,
at sites occupied by individuals of each species.  Aside from this interaction,
the two populations evolve independently.  We found that, as one would expect, the symbiotic interaction
favors survival of a mixed population, in that the critical reproduction rate $\lambda_c$ decreases
as we reduce $\mu$ \cite{scp}.

Apart from its interest as an elementary model of symbiosis, the critical
behavior of the two-species symbiotic CP (2SCP) is interesting for the study of nonequilibrium
universality classes. Extinction represents an absorbing state, a frozen state with no
fluctuations \cite{marro,henkel,odor07,hinrichsen,odor04}.
Absorbing-state phase transitions have been a topic of much interest in recent decades. In addition to
their connection with
population dynamics, they appear in a wide variety of problems, such as heterogeneous catalysis \cite{zgb},
interface growth \cite{tang}, and epidemics \cite{bart},  and have been shown to underlie
self-organized criticality \cite{vdmz,bjp}. Recent experimental realizations in the
context of spatio-temporal chaos in liquid crystal
electroconvection \cite{take07}, driven suspensions \cite{pine} and superconducting vortices \cite{okuma}
have heightened interest in such transitions. In this context, in \cite{scp} we employed extensive
simulations and field-theoretical arguments to show that the critical scaling of the 2SCP is consistent with
that of directed percolation (DP), which is known to describe the basic CP \cite{note1},
and is generic for absorbing-state phase transitions \cite{janssen,grassberger}.

In this work we examine some of the issues regarding the 2SCP left open in the original study \cite{scp}:
(1) Can mean-field predictions be improved on?  (2) What is the phase boundary for unequal creation rates?
(3) Does the model exhibit a discontinuous phase transition in two dimensions, for strong symbiosis, or
in the presence of diffusion?

The mean-field theory for the 2SCP \cite{scp}, at both one- and two-site levels,
predicts a discontinuous phase transition for strong symbiosis in any number of dimensions.
Discontinuous phase transitions to an absorbing state are not possible, however, in one-dimensional systems with
short-range interactions and free of boundary fields \cite{hinrichsen}.
We have indeed verified this general principle in simulations of the one-dimensional model.
The simulations reported in \cite{scp} did not reveal a discontinuous transition
in two dimensions ($d=2$) either.
In the present work we aim to provide a better theoretical understanding of the phase diagram of the 2SCP,
using exact quasistationary probability distributions for small systems, cluster approximations,
and simulations. In two dimensions, we extend the model to include diffusion (nearest-neighbor hopping)
of individuals.  While we find no evidence of a discontinuous transition without diffusion, it becomes
discontinuous for sufficiently small $\mu$ and large $D$.

The remainder of this paper is organized as follows. In
Sec. II we review the definition of the model and the mean-field analysis, and in
Sec. III present results of cluster approximations and quasistationary analysis.
Then, in Sec. IV we study the diffusive process. Sec. V is devoted to discussion and conclusions.

\section{Model}

To begin we review the definition of the two-species symbiotic contact process (2SCP) \cite{scp}.
We denote the variables for occupation of a site $i$ by species A and B as
$\sigma_i$ and $\eta_i$, respectively.  The possible states $(\sigma_i, \eta_i)$ of a given site
are $(0,0)$ (empty),
$(1,0)$ (occupied by species A only), $(0,1)$ (species B only), and $(1,1)$ (occupied by both species).
Birth of $A$ individuals, represented by the transitions
$(0,0) \to (1,0)$ and $(0,1) \to (1,1)$, occur at rate $\lambda_A r_A$, with $r_A$
the fraction of nearest neighbor sites (NNs) bearing a particle of species A.  Similarly, birth of $B$ individuals
[i.e., the transitions $(0,0) \to (0,1)$ and $(1,0) \to (1,1)$], occurs at rate $\lambda_B r_B$,
with $r_B$ the fraction of NNs bearing a particle of species B.
Death at singly occupied sites, $(1,0) \to (0,0)$ and $(0,1) \to (0,0)$, occurs at a rate of
unity, as in the basic CP.  The transitions $(1,1) \to (1,0)$ and $(1,1) \to (0,1)$,
corresponding to death at a doubly occupied site,
occur at rate $\mu$.
The set of transition rates defined above describes a pair of contact processes inhabiting the
same lattice.  If $\mu=1$ the two processes
evolve independently, but for $\mu < 1$ they interact {\em symbiotically} since
the annihilation rates are reduced at sites with both species present.

The phase diagram of the 2SCP exhibits four phases: (i) the fully active phase with nonzero populations
of both species; (ii) a partly active phase with only $A$ species; (iii) a partly active phase with only $B$ species;
(iv) the inactive phase in which both species are extinct.  The latter is absorbing while the
partly active phases represent absorbing subspaces of the dynamics. Extensive simulations on rings and on the
square lattice indicate that the critical behavior is compatible with the directed percolation (DP) universality class;
this conclusion is also supported by field-theoretic arguments \cite{scp}.

In \cite{scp}, we studied the model with symmetrical rates under exchange of species labels
A and B, i.e., with $\lambda_A=\lambda_B=\lambda$. We found that for $\mu < 1$
the transition from the fully
active to the absorbing phase occurs at some $\lambda_{c} (\mu) < \lambda_{c} (\mu=1)$, since the annihilation rate is reduced.
The effect of asymmetric creation rates is shown in Fig.~1 : if one of the species, for instance A,
has its creation rate below (above) $\lambda_c$, the transition occurs for a $\lambda_B$ above (below) $\lambda_c$.
 (The simulation algorithm is detailed in Sec. IV.)
The results for $d=2$ are qualitatively the same, as shown in Fig.~2.
Suppose we let $\lambda_A \to \infty$. Then all sites will bear an A particle, so that the dynamics
of species B is a contact process with death rate $\mu$.  It follows that the critical value of $\lambda_B$
is $\mu \, \lambda_c (\mu=1)$;  this determines the asymptotic form of the phase boundaries in Figs. 1 and 2.
The simulation data in Figs.~1 and 2 are obtained
by extrapolating moment ratio crossings \cite{moments}. The system sizes are $L=200$, 400, 800 and 1600 in one dimension,
and $L=40$, 80, 160 and 320 in two dimensions.

\begin{figure}[!hbt]
\includegraphics[clip,angle=0,width=0.8\hsize]{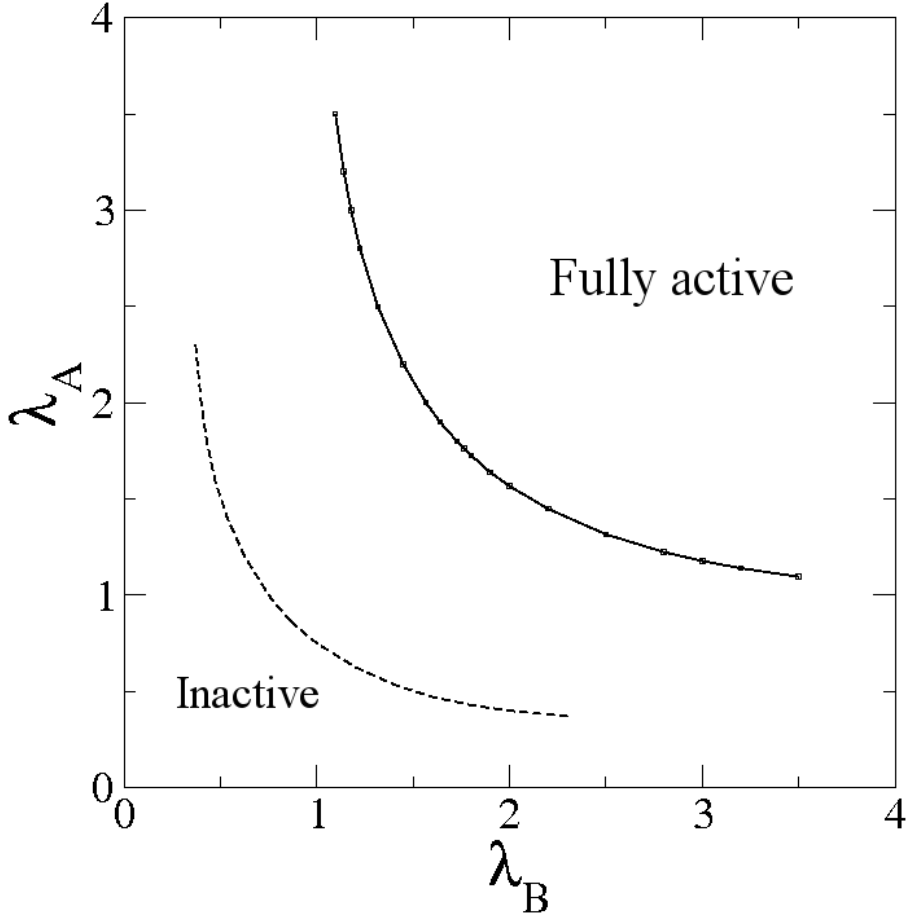}
\caption{\footnotesize{One-dimensional 2SCP: phase diagram in the
$\lambda_A$-$\lambda_B$ plane for $\mu = 0.25$, obtained via simulation (points) and mean field theory (dashed curve).
}}

\label{lalb1}
\end{figure}

\begin{figure}[!hbt]
\includegraphics[clip,angle=0,width=0.8\hsize]{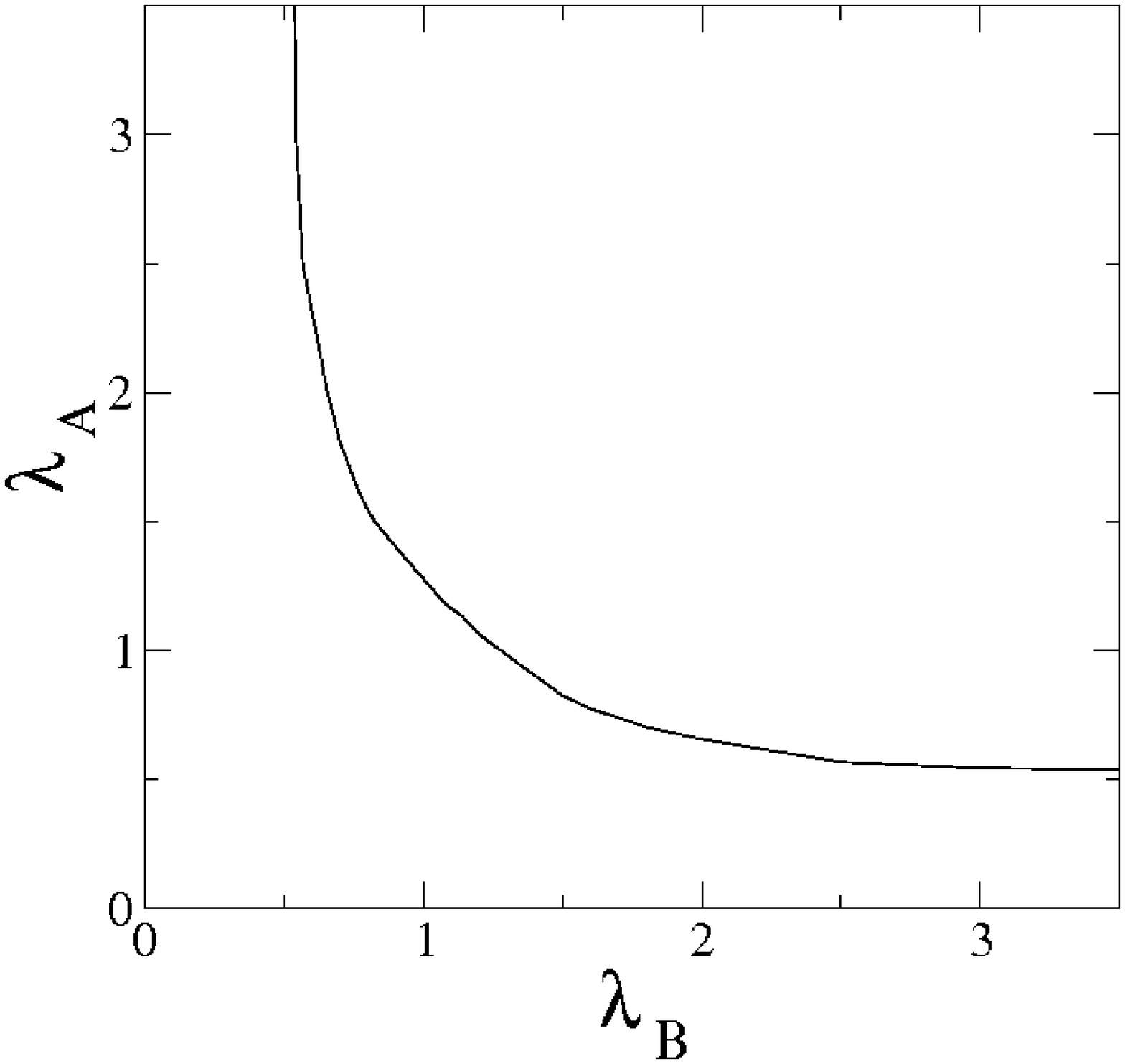}
\caption{\footnotesize{Two-dimensional 2SCP: phase diagram in the $\lambda_A$-$\lambda_B$ plane
for $\mu=0.25$, obtained via simulation.
}}

\label{lalb2}
\end{figure}

The basic mean-field theory (MFT)  (i.e., the one-site approximation),
for the 2SCP was derived in \cite{scp}.
Generalized to include different creation rates, $\lambda_A$ and $\lambda_B$, for the two species,
and diffusion (nearest-neighbor hopping) of both species at rate $D$, the MFT equations read:

\begin{eqnarray}
\frac{d p_0}{dt} &=& - (\lambda_A \rho_A + \lambda_B \rho_B)p_0 + p_A + p_B
+D[p_A \tilde{\rho}_A + p_B\tilde{\rho}_B - \rho p_0],
%\nonumber
\\
\frac{d p_A}{dt} &=&  \lambda_A p_0 \rho_A + \mu p_{AB} - (1 + \lambda_B \rho_B) p_A
+D[p_0\rho_A \!-\! p_A\rho_B +p_{AB}\tilde{\rho}_B \!-\! p_A\tilde{\rho}_A],
%\nonumber
\\
\frac{d p_B}{dt} &=&  \lambda_B p_0 \rho_B + \mu p_{AB} - (1 + \lambda_A \rho_A) p_B
+D[p_0\rho_B \!-\! p_B\rho_A +p_{AB}\tilde{\rho}_A \!-\! p_B\tilde{\rho}_B],
%\nonumber
\\
\frac{d p_{AB}}{dt} &=& \lambda_B p_A \rho_B + \lambda_A p_B \rho_A - 2 \mu p_{AB}
+D[p_A \rho_B + p_B\rho_A - p_{AB}(2-\rho)],
\label{pab}
\end{eqnarray}

\noindent where the probabilities for a given site to be
vacant, occupied by species A only, by species B only, and doubly occupied are denoted by $p_0$, $p_A$, $p_B$,
and $p_{AB}$, respectively, $\rho_A = p_A + p_{AB}$, and $\rho_B = p_B + p_{AB}$.
We have further defined $\rho = \rho_A + \rho_B$, $\tilde{\rho}_A = 1-\rho_A$ and $\tilde{\rho}_B = 1-\rho_B$.
If one species
is absent (for example, if $p_B = p_{AB} = 0$) this system reduces to the MFT
for the basic contact process, $\dot{p}_A = \lambda p_A (1-p_A) - p_A$, with a
critical point at $\lambda = 1$.  Under the effect of symbiosis we seek a
symmetric stationary solution, $p_A = p_B = p$, leading, for $D=0$, to

\begin{equation}
\overline{p} = \frac{\mu}{2 \lambda (1-\mu)} \left[ 2(1-\mu) - \lambda
+ \sqrt{\lambda^2 - 4\mu (1-\mu)} \right].
\label{pMFT}
\end{equation}
and
\begin{equation}
\overline{p}_{AB} = \frac{\lambda p^2}{\mu - \lambda p}
\label{pabmft}
\end{equation}

\noindent For $\mu \geq 1/2$, $p$ grows continuously from
zero at $\lambda=1$, marking the latter value as the critical point.
The activity grows linearly, $p \simeq [\mu/(2\mu -1)](\lambda-1)$, in this regime.
For $\mu < 1/2$, however, the expression is already positive for
$\lambda = \sqrt{4 \mu(1-\mu)} < 1$, and there is a {\it discontinuous}
transition at this point.

In the limit $D \to \infty$, we expect $p_{AB} = \rho_A \rho_B$, as is required by the condition that,
in this limit, a time-independent
solution requires that the coefficient of $D$ in Eq.~\ref{pab} be zero.

\section{Cluster approximations and quasistationary analysis}

As noted above, the discontinuous phase transition predicted by one- and two-site MFT is
impossible in one dimension.  Simulations in both one and two dimensions, covering a
broad range of $\mu$ values, yield no evidence of a discontinuous transition.
Here we attempt to develop more reliable theoretical descriptions, using cluster approximations
and quasistationary (QS) solutions of small systems, for the symmetric case, $\lambda_A = \lambda_B = \lambda$.
In the following analysis we set $D=0$, i.e., the non-diffusive limit of Eqs. 1 - 4.

\begin{figure}[!hbt]
\includegraphics[clip,angle=0,width=0.8\hsize]{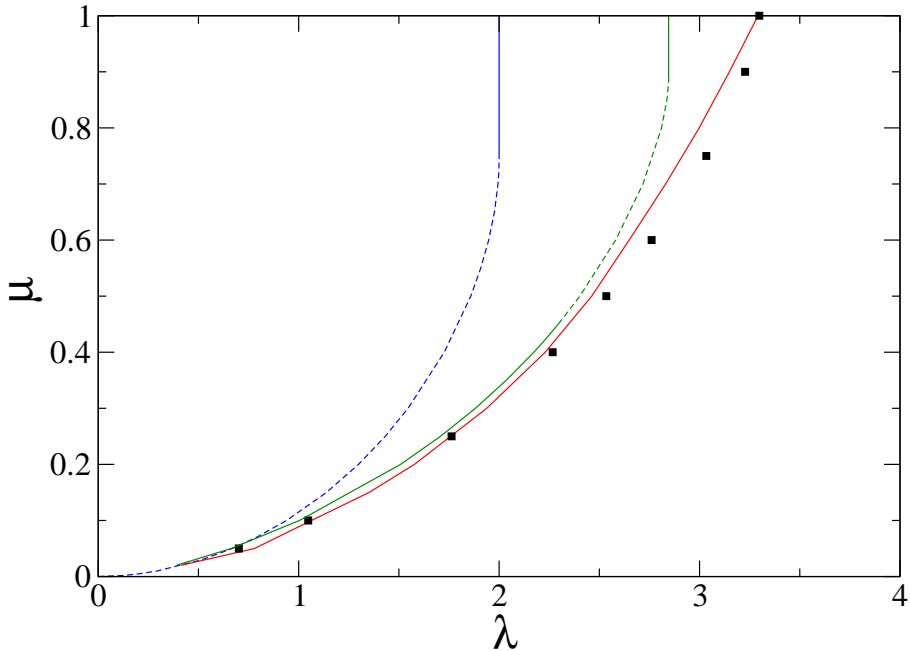}
\caption{\footnotesize{(Color online) 2SCP in one dimension: phase boundary in the $\lambda-\mu$ plane
as given by simulations (symbols). From left to right curves, phase boundary given by the 2-site (blue) and 6-site (green) cluster approximations,
and via analysis of the QS distribution (red). Straight (dashed) curves represent continuous (discontinuous) phase transitions.
}}
\label{lm1d}
\end{figure}

It is often the case that MFT predictions improve, both qualitatively and quantitatively,
as the cluster size used in the analysis is increased.  We therefore investigate MFT approximations
using clusters of up to six sites in one dimension, and clusters of four sites on the square lattice.
Following the usual procedure \cite{mftzgb,marro,benav}, we deduce a set of coupled, nonlinear
differential equations for
the cluster occupation probabilities, which are then integrated numerically to obtain the stationary solution.
As shown in Fig.~\ref{lm1d}, for the one-dimensional case, the prediction for the phase boundary in the
$\lambda-\mu$ plane does improve as we increase the cluster size from $n=2$ to $n=6$.
 The $n=2$ approximation correctly predicts a continuous phase transition for $\mu \geq 0.75$,
but on this range it yields $\lambda_c$ independent of $\mu$, contrary to simulations, which
show $\lambda_c$ varying smoothly with $\mu$. For $n=6$ the transition is predicted to be
continuous for $\mu < 0.45$, {\it discontinuous} for $0.45\leq\mu < 0.88 $, and again continuous for
$0.88 \leq \mu \leq 1$. (Note that on the latter interval $\lambda_c$ is
again independent of $\mu$). 
Thus the $n=6$ approximation exhibits the same qualitative problems as for $n=2$, despite the
overall improvement.
The four-site approximation on the square lattice, shown in Fig.~\ref{lm2d}, furnishes a
reasonable prediction for the phase boundary, but suffers from similar defects:
for $\mu < 0.66$ the transition is discontinuous, while for $\mu \geq 0.7$, $\lambda_c$ is independent of $\mu$.

In the context of absorbing-state phase transitions, we generally look to MFT as a guide to the
overall phase diagram, expecting the critical point to have the correct order of magnitude and,
perhaps more importantly, the nature (continuous or discontinuous) of the transition to
be predicted correctly.  The latter criterion is not always satisfied, however \cite{trpcr2009}.
In light of this, and in the hope of devising a more reliable approximation method that is still
relatively simple to apply, we consider analyses based on the quasistationary (QS) probability
distribution of small systems.  The QS distribution (or {\it Yaglom limit}, as it is known in the
probability literature), is the probability distribution at long times, conditioned on survival
of the process \cite{QSS}.  For the one-dimensional CP and allied models \cite{exact}, and an activated random
walker model \cite{sleepy}, finite-size scaling analysis of numerically exact QS results on a sequence of
lattice sizes yields good estimates for the critical point, exponents and moment ratios.
In the present case, with four states per site, attaining the sizes required for a precise analysis
appears to be very costly, computationally, and we shall merely attempt to obtain reasonable estimates
for the phase boundary $\lambda_{c}(\mu)$.

As described in detail in \cite{exact}, obtaining the QS distribution numerically requires (1) enumerating
all configurations on a lattice of a given size; (2) enumerating all transitions between
configurations, and their associated rates; and (3) using this information in an iterative procedure to generate
the QS distribution.  Once the latter is known, one may calculate properties such as the
order parameter or lifetime.  For small systems these quantities are smooth functions
of the control parameter and show no hint of the critical singularity.  It is known, however, that the
moment ratio $m(\lambda;L) \equiv \langle \rho^2 \rangle/\langle \rho \rangle^2$ exhibits crossings,
analogous to those of the Binder cumulant \cite{moments}.  (Here $\rho$ is the density of active sites.)
That is, defining $\lambda_\times (L)$ via the condition
$m[\lambda_\times (L);L] = m[\lambda_\times (L);L\!-\!1]$, the $\lambda_\times (L)$
converge to $\lambda_c$ as $L \to \infty$,
as follows from a scaling property of the order-parameter probability distribution.
Our procedure, therefore, is to calculate $m(\lambda;L)$ for a series of sizes $L$, locate the
crossings $\lambda_\times (L)$, and use them to estimate $\lambda_c$.

In one dimension we calculate $m(\lambda;L)$ for rings of size $L=6$ to 11.  We treat configurations
with only one species as absorbing, as well as, naturally, the configuration devoid of any individuals.
To estimate $\lambda_c$
we perform a quadratic fit to $\lambda_\times (L)$ as a function of $L^{-\gamma}$, using $\gamma$ in the range 1-3.
(The precise value of $\gamma$ is chosen so as to render the plot of $\lambda(L)$ versus $L^{-\gamma}$
as close to linear as possible.)  Similar estimates for $\lambda_c$ are obtained using the Bulirsch-Stoer
procedure \cite{BS}.  As is evident in Fig.~\ref{lm1d}, the resulting phase boundary is in good accord with
simulation, predicting $\lambda_{c,\mu}$ with an accuracy of 10\% or better. (The simulation data in Fig.~3 are obtained
by extrapolating moment ratio crossings \cite{moments} for system sizes $L=200$, 400, 800 and 1600).  The extrapolated value of
$m$ at the crossings is not particularly good (for $\mu=1$ we find $m_c = 1.110$, compared with the
best estimate of 1.1736(1) \cite{moments}). Although we expect that this would improve using larger systems,
our objective here is to find a relatively fast and simple method to predict the phase boundary.
(The cpu time required to converge to the QS distribution is comparable to that required to
integrate the equations numerically in the $n=6$ cluster approximation.)

\begin{figure}[!hbt]
\includegraphics[clip,angle=0,width=0.8\hsize]{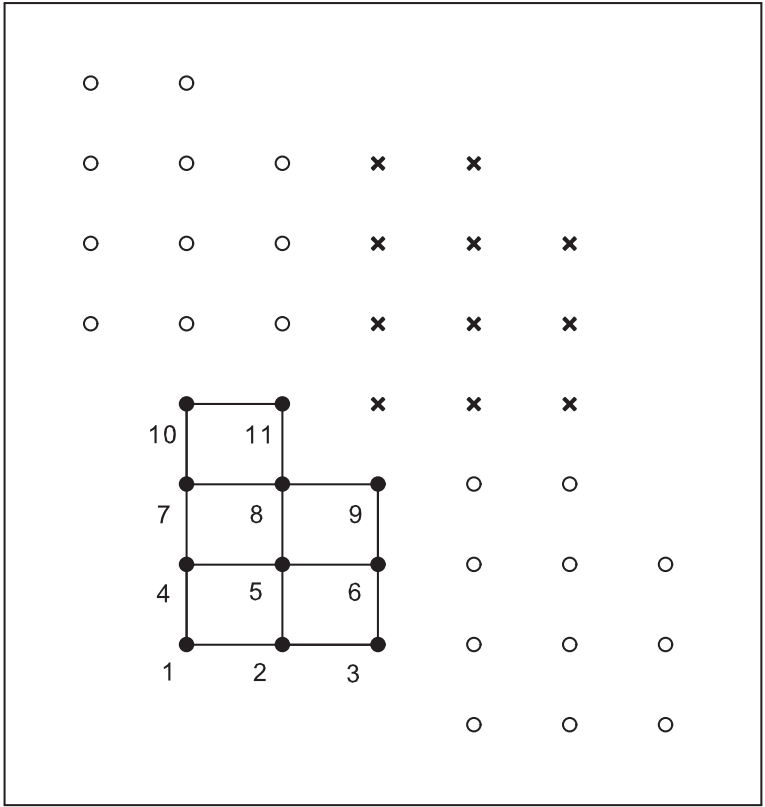}
%\vspace{1cm}

\caption{\footnotesize{Eleven-site cluster used in the QS analysis on the square lattice.  Copies
are placed so as to tile the plane; the tiling defines the neighbors for boundary sites, so that, for
example, the neighbors of site 1 are sites 2, 4, 9 and 11.
}}
\label{11site}
\end{figure}

\begin{figure}[!hbt]
\includegraphics[clip,angle=0,width=0.8\hsize]{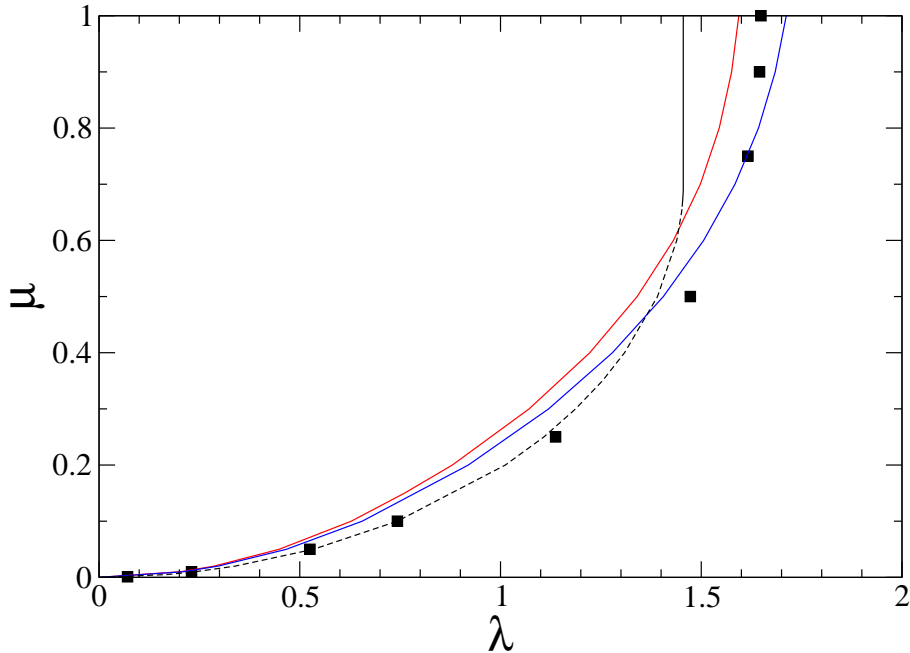}
\caption{\footnotesize{(Color online) Two-species CP on square lattice:
Phase boundary in the $\lambda-\mu$ plane as given by simulations (symbols),
by the 4-site cluster approximation (black curve), showing continuous (straight) and discontinuous (dashed) phase transitions, and by the quasi-stationary distributions for clusters of
11, red (upper) curve and 12 sites, blue (bottom) curve.
}}
\label{lm2d}
\end{figure}

To apply the QS method to the two-dimensional 2SCP, we devised an algorithm that enumerates
configurations and transitions for a general graph of $N$ vertices; the graph structure is
specified by the set of bonds ${\cal B} = \{(i_1,j_1), (i_2,j_2), ..., (i_m,j_m)\}$ linking
pairs of vertices $i_k$ and $j_k$.  To represent a portion of the square lattice, with periodic
boundaries, each vertex must be linked to four others.  This can be achieved rather naturally
for a square ($m \times m$) or rectangle ($m \times (m+1)$); for other values of $N$ we use a
cluster close to a square, and define the bonds required for periodicity by tiling the plane
with this cluster, as shown in Fig.~\ref{11site}.

We study clusters of 8 to 12 sites on the square lattice.
For $N=12$, there are about 1.7 $\times 10^7$ configurations and about 3.9 $\times 10^8$ transitions; restrictions
of computer time and storage prevent us from going beyond this size.
The crossings of $m$ between successive sizes do not yield useful predictions for $\lambda_c$
in this case.  Evidently, the linear extent of the clusters is too small to probe the scaling regime.
We instead derive estimates for the critical point by locating the maximum of $d\rho/d\lambda$, since
in the infinite-size limit, this derivative (taken from the left) diverges at the critical point.
The resulting predictions, for clusters of 11 and 12 sites, are compared with simulation in Fig.~\ref{lm2d},
showing that the QS analysis provides a semiquantitative prediction for $\lambda_c$, and captures
the shape of the phase boundary.  (The simulation data in Fig. 5 are obtained
by extrapolating moment ratio crossings \cite{moments} for system with linear sizes $L=40$, 80, 160 and 320).
This analysis suggests that the phase transition is continuous
(as found in simulation) since the QS probability distribution is unimodal in all cases.

\section{The Diffusive SCP}

Although the one-site MFT predicts a discontinuous phase transition in the 2SCP in any number of dimensions,
such a transition is not possible in one-dimensional systems with short-range interactions and free of
boundary fields \cite{hinrichsen}.
In one dimension the active-absorbing transition should be continuous, as we have
indeed verified in simulations. In two dimensions ($d=2$), previous studies did not reveal any
evidence  for a discontinuous transition.  These studies did not, however, include diffusion,
which is expected to facilitate the appearance of discontinuous transitions.
Here we study the 2SCP with diffusion on the square lattice.

We modify the process so that, in addition to creation and death, each individual
can hop to one of its NN sites at
rate $D$. In the simulation algorithm for the diffusive 2SCP, we maintain two lists, one of singly
and another of doubly occupied sites.  Let $N_s$ and $N_d$ denote, respectively, the numbers of such
sites, so that $N_p = N_s + 2 N_d$ is the total number of individuals.  The total rate of (attempted)
transitions is $\lambda N_p + N_s + 2\mu N_d + D N_p\equiv 1/\Delta t$, where $\Delta t$ is the
time increment associated with a given step in the simulation.

At each such step, we choose
among the events: (1) creation attempt by an isolated individual, with probability $\lambda N_s \Delta t$;
(2) creation attempt by an individual at a doubly occupied site, with probability $2 \lambda N_d \Delta t$;
(3) death of an isolated individual, with probability $N_s \Delta t$; (4)
death of an individual at a doubly occupied site, with probability $2 \mu N_d$ and (5) diffusion of an
individual, with probability $D N_p \Delta t$.

Once the event type is selected a site $i$ is randomly chosen from the appropriate list.
Creation occurs at a site $j$, a randomly chosen first-neighbor of site $i$, if $j$ is not already
occupied by an individual of the species to be created.
If site $i$ is doubly occupied, the species of the daughter (in a creation event)
is chosen to be A or B with equal probability.  Similarly, in an annihilation event at a
doubly-occupied site, the species to be removed is chosen at random.

For the SCP with diffusion, we performed QS simulations \cite{qssim,qssim2} for systems of linear sizes up to $L= 100$, with each run lasting $10^8$
time units.
Averages are taken in the QS regime, after discarding an initial transient which depends on the system size and
diffusion rate used.

\begin{figure}[!hbt]
\includegraphics[clip,angle=0,width=0.8\hsize]{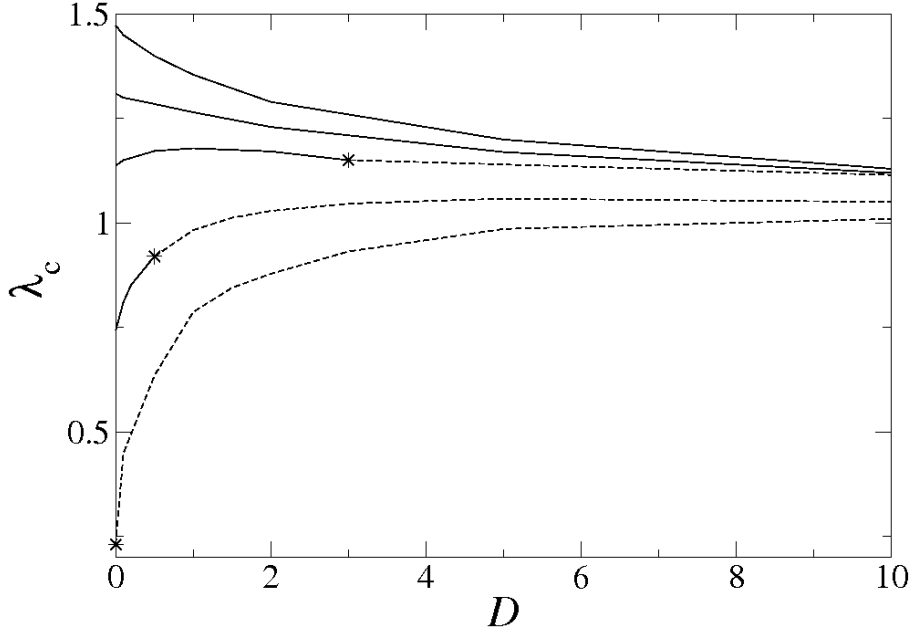}
\caption{\footnotesize{(Color online) Diffusive 2SCP on the square lattice:  critical creation rate $\lambda_c$
versus diffusion rate $D$, for $\mu=0.01, 0.1, 0.25, 0.35$ and $0.5$, from bottom to top. Solid (dashed) lines represent
continuous (discontinuous) phase transitions. The star represents the tricritical point for $\mu=0.01,0.1$ and $0.25$. System size: $L=100$.
}}
\label{scpD}
\end{figure}

\begin{figure}[!hbt]
\includegraphics[clip,angle=0,width=0.8\hsize]{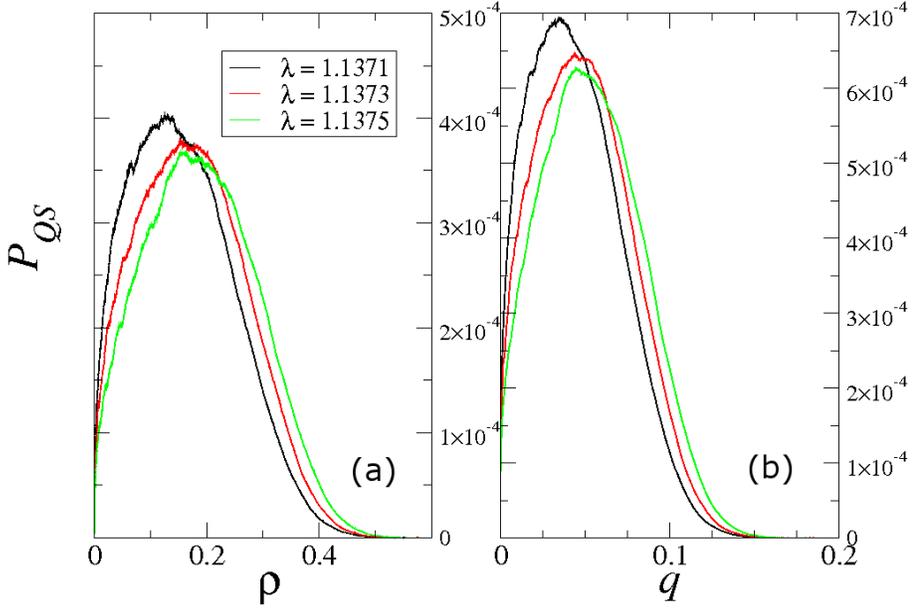}
\caption{\footnotesize{(Color online) 2SCP on square lattice:
QS probability distributions of $\rho$ (a)  and $q$ (b),
for $\mu=0.25$, $D=0$, and (left to right) $\lambda=1.1371$,
$\lambda=1.1373$ and  $\lambda=1.1375$. System size $L=100$.}}
\label{mu25d0}
\end{figure}

\begin{figure}[!hbt]
\includegraphics[clip,angle=0,width=0.8\hsize]{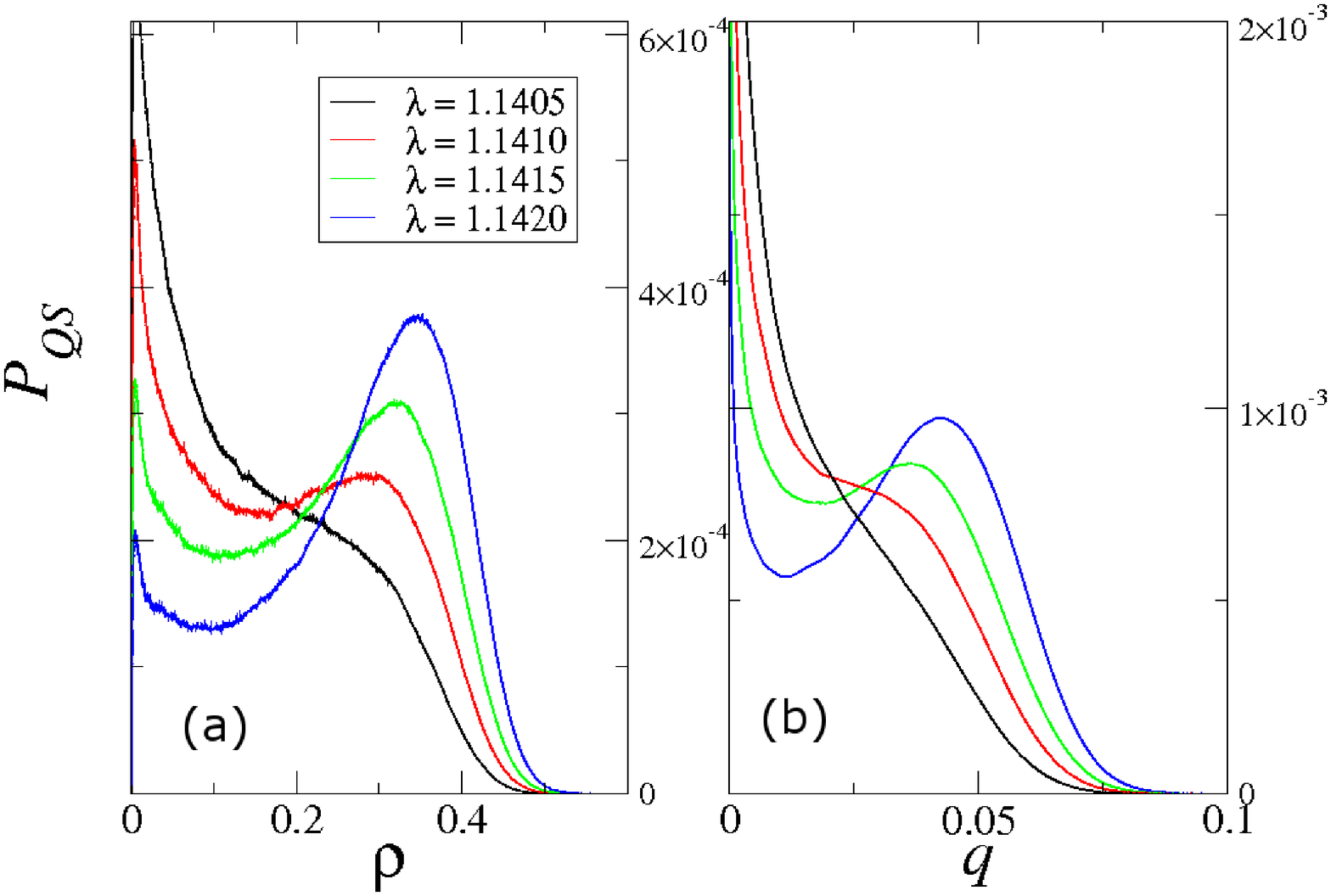}
\caption{\footnotesize{(Color online)  2SCP on the square lattice: QS probability distributions
of $\rho$ (a) and $q$ (b),
for $\mu=0.25$, $D=5.0$, and $\lambda=1.1405$,
$\lambda=1.1410$, $\lambda=1.1415$ and $\lambda=1.1420$. System size $L=100$.
}}
\label{mu25d5}
\end{figure}

Figure \ref{scpD} shows that with increasing diffusion rate, the critical creation rate $\lambda_c$ tends
to unity, the value predicted by simple mean-field theory.
(The increase in $\lambda_c$ in the small-$D$ regime reflects the elimination
symbiotic A-B pairs due to diffusion.)
In Fig.~\ref{mu25d0} we plot near-critical quasistationary probability distributions of single
individuals, $\rho$, and of doubly occupied sites, $q$, for $\mu=0.25$ and $D=0$.
The distributions are unimodal, showing that the transition is continuous.
We verify that in the absence of diffusion, the absorbing phase transition is always continuous, regardless the
value of $\mu$.
For diffusion rates considerably in excess of unity, we observe a discontinuous
transition for certain values of $\mu$.
An example of bimodal QS probability distributions, signaling a discontinuous transition,
is shown in Fig.~\ref{mu25d5}, for $D=5.0$.\\

\begin{figure}[!hbt]
\includegraphics[clip,angle=0,width=0.8\hsize]{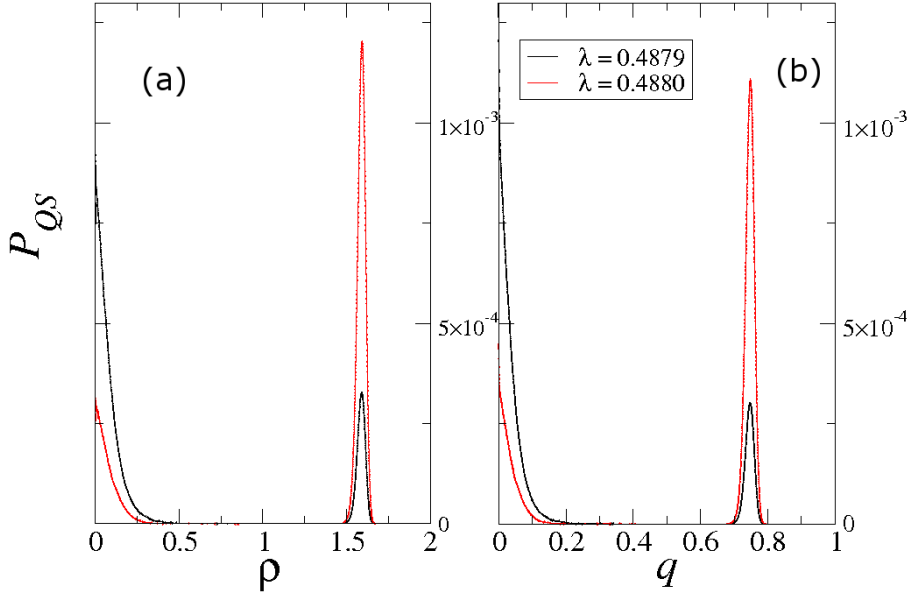}
\caption{\footnotesize{(Color online)  2SCP on the square lattice: QS probability distributions
of $\rho$ (a) and $q$ (b),
for $\mu=0.01$, $D=0.1$, and $\lambda=0.4879$ (black curves) and $\lambda=0.4880$ in red (gray). System size $L=100$.
}}
\label{mu01d.1}
\end{figure}

\begin{figure}[!hbt]
\includegraphics[clip,angle=0,width=0.95\hsize]{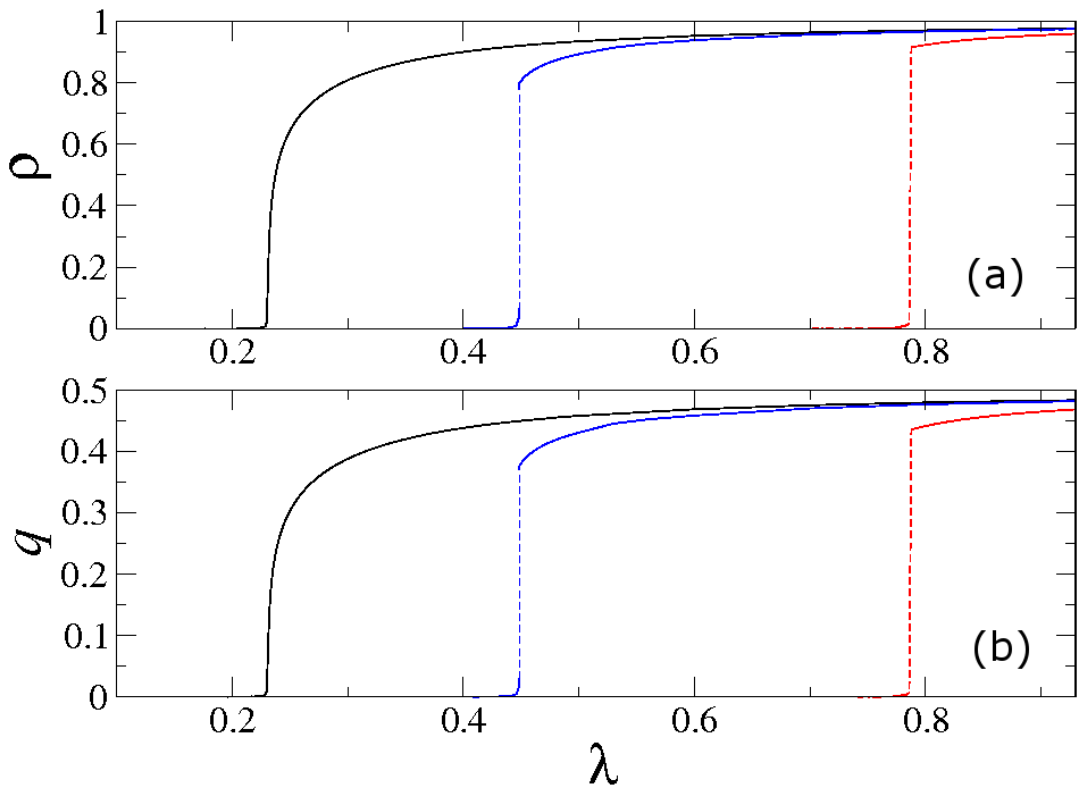}
\caption{\footnotesize{(Color online)  QS densities of $\rho$ (a) and $q$ (b),
for $\mu=0.01$ and $D=0$, $0.1$ and $1.0$, from left to right. System size $L=100$.
}}
\label{rqd01}
\end{figure}

\begin{figure}[!hbt]
\includegraphics[clip,angle=0,width=0.8\hsize]{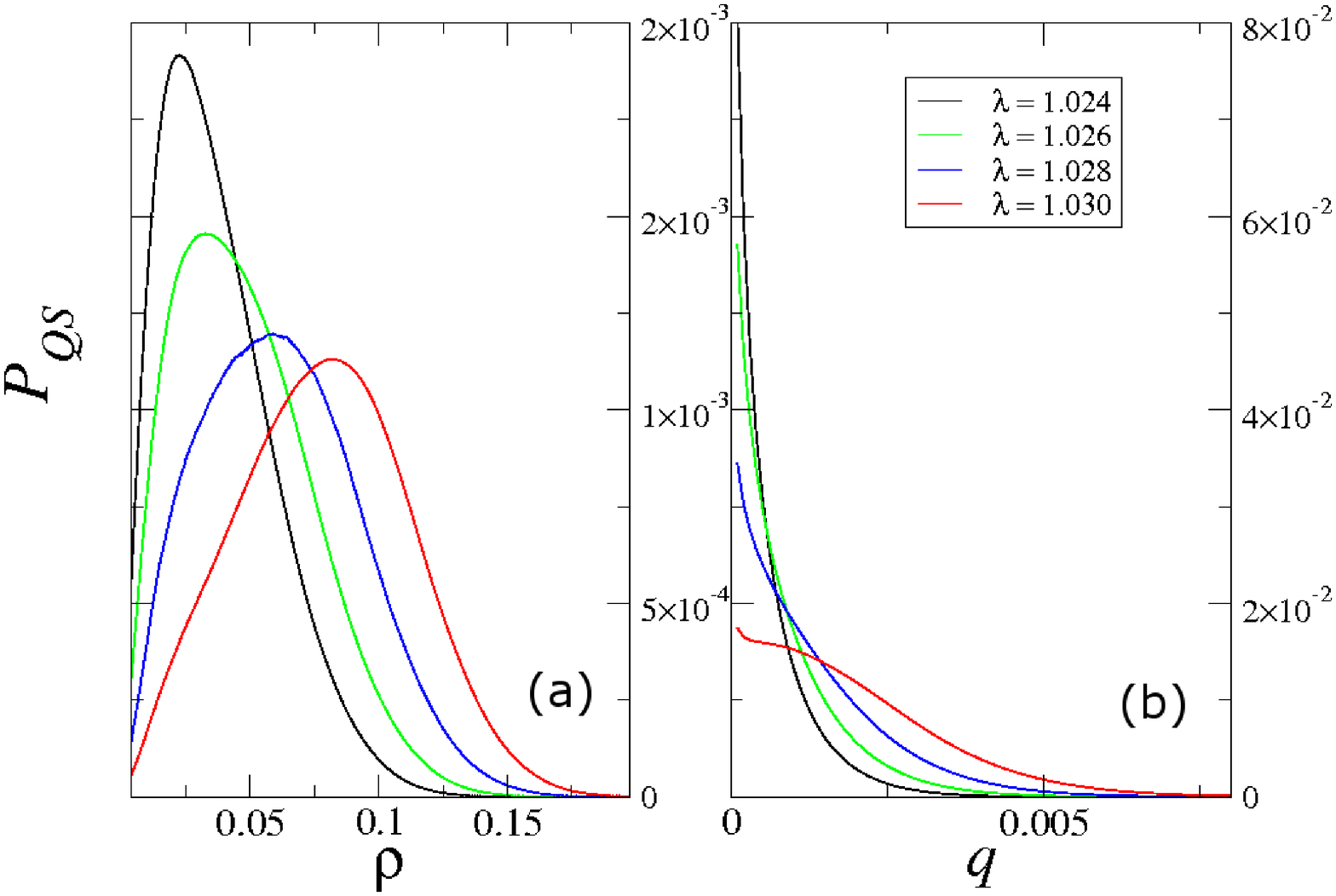}
\caption{\footnotesize{(Color online)  2SCP on the square lattice: QS probability distributions
of $\rho$ (a) and $q$ (b),
for $\mu=0.25$, $D=100$, and $\lambda=1.024$ (black curves),$\lambda=1.026$ (green) ,$\lambda=1.028$ (blue)  and $\lambda=1.030$ (red). System size $L=100$.
}}
\label{mu25d100}
\end{figure}

The mechanism by which diffusion gives rise to a discontinuous transition can be understood as follows.
Under strong symbiosis ($\mu$ close to zero), only doubly occupied sites are observed near the
critical point, in the absence of diffusion.  Since the transition is continuous in this case, the
overall density is very low near the critical point.
In the presence of diffusion, pairs tend to be destroyed; the resulting isolated individuals then rapidly die.
Thus diffusion renders low-density active states inviable.  Under moderate diffusion, a finite density is
required to maintain a significant concentration of doubly occupied sites, and thereby maintain activity.
Hence the population density jumps from zero to a finite value at the transition.
For small $\mu$ we observe a discontinuous phase transition even  for
small values of the diffusion rate, as shown in Figs.~\ref{mu01d.1} and \ref{rqd01}.

\begin{figure}[!hbt]
\includegraphics[clip,angle=0,width=0.95\hsize]{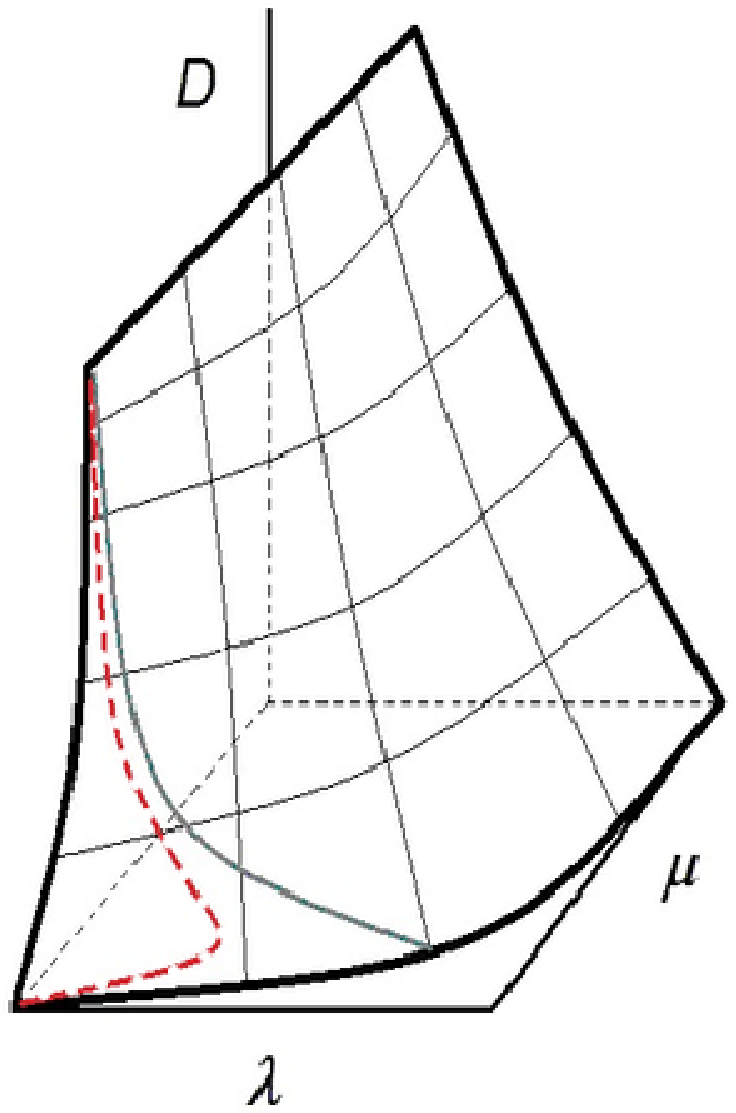}
\caption{\footnotesize{(Color online)  Schematic of the critical surface in $\lambda-\mu-D$ space,
showing the critical surface and the tricritical line on this surface, as predicted by MFT, in solid (grey) line,
and observed in simulations on the square lattice, in dashed (red) line.
}}
\label{crsurf}
\end{figure}

Although we have verified that the phase transition is
discontinuous for small $\mu$ and moderate diffusion rates $D$,
increasing $D$ further, the transition becomes continuous again.
In the limit $D \to \infty$, we expect mean field-like behavior, with the effects of diffusion suppressing
the clustering which permits symbiosis.  In this limit, the one-site MFT predicts
a {\it continuous} phase transition, with $\lambda_c = 1$, for any value of $\mu$.
Reversion to a continuous transition under rapid diffusion ($D=100$, $\mu=0.25$) is evident in Fig.~\ref{mu25d100}:
the QS probability distributions are again unimodal.
At criticality, fewer than $4\%$ of the individuals are located at
doubly occupied sites for $D=100$, in comparison with $25\%$ for $D=5$.

In the three-dimensional parameter space
space of $\lambda$, $\mu$, and $D$, there is a critical surface separating the active and absorbing phases.
On this surface, a {\it tricritical line} separates regions exhibiting continuous and discontinuous phase
transitions (see Fig.~\ref{crsurf}).
 The mean-field theory of Eqs.~(1)-(4) yields a tricritical line that begins at $\lambda=1$,
$\mu=1/2$ (for $D=0$), and then tends, for increasing $D$, to ever smaller values of $\mu$ (asymptotically,
$\mu = 1/D$, with $\lambda=1$ all the while).  Simulations show a somewhat different picture, with the
tricritical line approaching the point $\lambda=\mu=D=0$, and then curving toward larger $\mu$ and $\lambda$
values for small but nonzero $D$, before doubling back towards $\mu=0$, as shown in Fig~\ref{crsurf}.  This means that for a
given, nonzero value of $\mu$, the transition is discontinuous (if at all), only within a restricted
range of $D$ values.
For example, our simulations reveal that for $\mu=0.25$,
the transition is discontinuous for $3 < D < 10$, but becomes
continuous for $D \geq 100$.
We defer a full mapping of the tricritical line to future work.

\section{Conclusions}

We present a detailed study of the phase diagram of the symbiotic contact process, using simulation,
cluster approximations, and exact (numerical) quasistationary distributions of small systems.
We study the effect of asymmetric creation rates and of diffusion of individuals.
Exact quasistationary distributions and cluster approximations provide fair predictions for the
phase boundary in the symmetric case.
In simulations, the phase transition is always found to be continuous
in one dimension, but in two dimensions we observe a discontinuous phase
transition when symbiosis is strong ($\mu\to 0$),
in the presence of moderate diffusion.  For $D \to \infty$ the transition is again continuous.

Although the model studied here is much too simple to apply to real
ecosystems, our results raise the possibility of catastrophic (discontinuous) collapse of strongly
symbiotic interspecies alliances under increasingly adverse conditions, even if the change is gradual.
Possible extensions of this work include precise determination of the tricritical line
for the diffusive process, as well as the design of more precise theoretical approaches
for two-dimensional problems. The latter task assumes even greater significance when one observes
that despite the simplicity of the model, the full parameter space, including distinct
reproduction, death, and diffusion rates for each species, is far too vast to be mapped out
via simulation alone.
Finally,
the possibility of discontinuous phase transitions in more complex models of symbiosis
merits investigation.

\vspace{3cm}

\noindent{\bf Acknowledgments}

This work was supported by CNPq and FAPEMIG, Brazil.

\bibliographystyle{apsrev}

\begin{thebibliography}{100}

\bibitem{harris-CP}
        T.~E. Harris,
        Ann. Probab., {\bf 2}, 969 (1974).

\bibitem{durrett}
        R. Durrett,
        SIAM Rev. {\bf 41}, 677 (1994).

\bibitem{jansen}
        H. Janssen,
        J. Stat. Phys. 103, 801 (2001).

\bibitem{iwata}
        S. Iwata, K. Kobayashi, S. Higa, J. Yoshimura and K. Tainaka,
        Ecol. Modelling {\bf 222},  2042 (2011).

\bibitem{multisp}
        D.~C. Markham, M.~J. Simpson, P.~K. Maini, E.~A. Gaffney and  R.~E. Baker,
        Phys. Rev. E {\bf 88}, 052713 (2013).

\bibitem{parasites} S. J. Court , R.A. Blythe and R. J. Allen, Europhys. Lett. {\bf 101}, 50001 (2013).

\bibitem{competing}  T. B. Pedro, M. M. Szortyka and W. Figueiredo, J. Stat. Mech. {\bf 2014} P05016 (2014).

\bibitem{tauber} U. Dobramysl and U. C. Tauber, Phys. Rev. Lett. {\bf 110}, 048105 (2013).

\bibitem{tubay} J. M. Tubay et.al, Sci. Reports  {\bf 3}, 2835 (2013).

\bibitem{weitz} J. S. Weitz and D. H. Rothman, J. Theor. Biol. 225, 205 (2003).

\bibitem{munoz} M. Cencini, S. Pigolotti, M. A.  Mu\~noz, PloS One 7 (6), e38232 (2012).

\bibitem{boucher} D. Boucher, {\it The Biology of Mutualism: Ecology and Evolution} (Oxford University, New York, 1988).

\bibitem{douglas} A. E. Douglas, {\it Symbiotic Interactions} (Oxford University, Oxford, 1994).

\bibitem{sapp} J. Sapp, {\it Evolution by Association: A History of Symbiosis} (Oxford University, Oxford,1994).

\bibitem{paracer} S. Paracer and V. Ahmadjian, {\it Symbiosis: An introduction
to biological associations} (Oxford University Press, Oxford, 2nd ed., 2000).

\bibitem{rockwood} L. L. Rockwood, {\it Introduction to Population Ecology} (Blackwell Publishing, Malden, 2006).

\bibitem{yukalov} V. I. Yukalov, E. P. Yukalova and D. Sornette, Physica D {\bf 241}, 1270 (2012).

\bibitem{discrete} R. Durrett and S. Levin, Theor. Pop. Biol. {\bf 46}, 363 (1994).

\bibitem{korolev} K. Korolev and D. R. Nelson, Phys. Rev. Lett. {\bf 107}, 088103  (2011).

\bibitem{asta} L. Dall'Asta, F. Caccioli, and D. Begh\'e, Europhys. Lett.
{\bf 101}, 18003 (2013).

\bibitem{pigolotti} S. Pigolotti, R. Benzi, P. Perlekar, M. H. Jensen, F. Toschi and D. R. Nelson, Theor. Pop. Biol. {\bf 84}, 72 (2013).

\bibitem{lav} M. O. Lavrentovich and D. R. Nelson, Phys. Rev. Lett. {\bf 112}, 138102 (2014).

\bibitem{scp}
        M.~M. de Oliveira, R.~V. dos Santos and R. Dickman,
        Phys. Rev. E {\bf 86}, 011121 (2012).

\bibitem{marro}
        J. Marro and R. Dickman,
        {\it Nonequilibrium Phase Transitions in Lattice Models}
        (Cambridge University Press, Cambridge, 1999).

\bibitem{odor07}
        G. \'Odor,
        {\it Universality In Nonequilibrium Lattice Systems: Theoretical Foundations}
        (World Scientific,Singapore, 2007)

\bibitem{henkel}
        M. Henkel, H. Hinrichsen and S. Lubeck,
        {\it Non-Equilibrium Phase Transitions Volume I: Absorbing Phase Transitions}
        (Springer-Verlag, The Netherlands, 2008).

\bibitem{hinrichsen}
        H. Hinrichsen,
        Adv. Phys. {\bf 49}, 815 (2000).

\bibitem{odor04}
        G. \'Odor,
        Rev. Mod. Phys {\bf 76},  663 (2004).

\bibitem{zgb} R. M. Ziff, E. Gulari, and Y. Barshad, Phys. Rev. Lett. 56, 2553 (1986).

\bibitem{tang} L. H. Tang and H. Leschhorn, Phys. Rev. A 45, R8309(1992).

\bibitem{bart}
        M.~S. Bartlett,
        {\it Stochastic Population Models in Ecology and Epidemiology}
        (Methuen, London, 1960).

\bibitem{vdmz}
        A. Vespignani, R. Dickman, M. A. Mu\~noz, and S. Zapperi,
        Phys. Rev. Lett. {\bf 81}, 5676 (1998).

\bibitem{bjp}
        R. Dickman, M.~A. Mu\~noz, A. Vespignani, and S. Zapperi,
        Braz. J. Phys. {\bf 30}, 27 (2000).

\bibitem{take07}
        K.~A. Takeuchi, M. Kuroda, H. Chat\'e, and M. Sano,
        Phys. Rev. Lett. {\bf 99}, 234503 (2007).

\bibitem{pine}
        L. Cort\'e, P. M. Chaikin, J. P. Gollub, and D. J. Pine,
        Nature Physics {\bf 4}, 420 (2008).

\bibitem{okuma} S. Okuma, Y. Tsugawa, and A. Motohashi,
        Phys. Rev. B{\bf 83}, 012503 (2011).

\bibitem{cpsl}
        M. M. de Oliveira and R. Dickman,
        Phys. Rev. E {\bf 84}, 011125 (2011)

\bibitem{note1}
        Similar conclusions apply
        to a related model, the CP with creation at second-neighbor
        sites, in which each species inhabits a distinct sublattice \cite{cpsl} with enhanced survival at first
        neighbors.


\bibitem{janssen}
        H.~K. Janssen,
        Z. Phys. B {\bf 42}, 151 (1981).

\bibitem{grassberger}
        P. Grassberger,
        Z. Phys. B {\bf 47}, 365 (1982).

\bibitem{mftzgb}
        R. Dickman,
        Phys. Rev. A {\bf 34}, 4246 (1986).

\bibitem{benav}
        D. {ben-Avraham} and J. K\"ohler,
        Phys. Rev. {\bf A} 45, 8358 (1992).

\bibitem{trpcr2009}
        A well known example is the triplet-creation model; see
        G. \'Odor and R. Dickman,
        J. Stat. Mech. {\bf 2009} P08024, and references therein.

\bibitem{QSS}
        See R. Dickman and R. Vidigal, J. Phys. A {\bf 35},
        1147 (2002), and references therein.

\bibitem{exact}
        R. Dickman,
        Phys. Rev. E {\bf 73}, 036131 (2006).

\bibitem{sleepy}
        J.~C. Mansur Filho and R. Dickman,
        J. Stat. Mech. {\bf 2011}, P05029 (2011).


\bibitem{moments}
        R. Dickman and J. {Kamphorst Leal da Silva},
        Phys. Rev. E {\bf 58}, 4266 (1998).

\bibitem{BS}
        M. Henkel and G. Sch\"utz,
        J. Phys. A {\bf 21}, 2617 (1988).

\bibitem{qssim}
       M.~M. de Oliveira and R. Dickman,
        Phys. Rev. E {\bf 71}, 016129 (2005);
        R. Dickman and M. M. de Oliveira,
        Physica A {\bf 357}, 134 (2005).

\bibitem{qssim2}
        M.~M. de Oliveira and R. Dickman,
        Braz. J. Phys. {\bf 36}, 685 (2006).
\end{thebibliography}

\newpage

\end{document}